\documentclass{pasj00}
\def\ASPConf#1#2{ASP Conf. Ser. #1, #2}
\def\PublisherASP{San Francisco: ASP}

\begin{document}
\SetRunningHead{Author(s) in page-head}{Running Head}
\Received{2009/03/27}
\Accepted{2009/06/08}

\title{Optical and Near-Infrared Photometric Observation\\ during
the Superoutburst of the WZ~Sge-Type Dwarf Nova, \\V455 Andromedae}

\author{
Risako \textsc{Matsui}\altaffilmark{1},
Makoto \textsc{Uemura}\altaffilmark{2},
Akira \textsc{Arai}\altaffilmark{1},
Mahito \textsc{Sasada}\altaffilmark{1},
Takashi \textsc{Ohsugi}\altaffilmark{2},\\
Takuya \textsc{Yamashita}\altaffilmark{3},
Koji \textsc{Kawabata}\altaffilmark{2},
Yasushi \textsc{Fukazawa}\altaffilmark{1},
Tsumefumi \textsc{Mizuno}\altaffilmark{1},\\
Hideaki \textsc{Katagiri}\altaffilmark{1},
Hiromitsu \textsc{Takahashi}\altaffilmark{2},
Shuji \textsc{Sato}\altaffilmark{4},
Masaru \textsc{Kino}\altaffilmark{4},
Michitoshi \textsc{Yoshida}\altaffilmark{5},
Yasuhiro \textsc{Shimizu}\altaffilmark{5},
Shogo \textsc{Nagayama}\altaffilmark{5},
Kenshi \textsc{Yanagisawa}\altaffilmark{5},
Hiroyuki \textsc{Toda}\altaffilmark{5},\\
Kiichi \textsc{Okita}\altaffilmark{5}, and
Nobuyuki \textsc{Kawai}\altaffilmark{6}}

\altaffiltext{1}{Department of Physical Science, Hiroshima University,
Kagamiyama 1-3-1, \\Higashi-Hiroshima 739-8526}
\altaffiltext{2}{Astrophysical Science Center, Hiroshima University, Kagamiyama
1-3-1, \\Higashi-Hiroshima 739-8526}
\altaffiltext{3}{National Astronomical Observatory of Japan, Osawa 2-21-1,\\ Mitaka,
Tokyo 181-8588, Japan}
\altaffiltext{4}{Department of Physics, Nagoya University, Furo-cho,
Chikusa-ku, Nagoya 464-8602}
\altaffiltext{5}{Okayama Astrophysical Observatory, National
Astronomical Observatory of Japan, \\Kamogata Okayama 719-0232}
\altaffiltext{6}{Department of Physics, Tokyo Institute of Technology,
2-12-1 Ookayama, Meguro-ku, Tokyo 152-8551, Japan}


\maketitle
      
\begin{abstract}
We report on optical and infrared photometric observations of a WZ
 Sge-type dwarf nova, V455 And during a superoutburst in 2007.
These observations were performed with the KANATA ($V$, $J$,
 and $K_s$ bands) and MITSuME ($g'$, $Rc$, and $Ic$ bands) telescopes.
Our 6-band simultaneous observations allowed us to investigate the
 temporal variation of the temperature and the size of the emitting region
 associated with the superoutburst and short-term modulations, such as
 early and ordinary superhumps.
A hot ($>11000$~K) accretion disk suddenly disappeared when the
 superoutburst finished, while blackbody emission, probably from
 the disk, still remained dominant in the optical region with a
 moderately high temperature ($\sim 8000$~K).
This indicates that a substantial amount of gas was stored in the disk
 even after the outburst.
This remnant matter may be a sign of an expected mass-reservoir which
 can trigger echo outbursts observed in several WZ~Sge stars.
The color variation associated with superhumps indicates that viscous
 heating in a superhump source stopped on the way to the superhump maximum, and a subsequent
 expansion of a low-temperature region made the maximum.
The color variation of early superhumps was totally different from
 that of superhumps: the object was bluest at the early superhump
 minimum.
The temperature of the early superhump light source was lower than that
 of an underlying component, indicating that the early superhump light source was a
 vertically expanded low-temperature region at the outermost part of the
 disk.
\end{abstract}

\section{Introduction}
Cataclysmic variables are
semi-detached binary systems consisting of a primary white dwarf and a
secondary red star.
A Roche-lobe-filling red star loses mass through the inner Lagrangian
point and the white dwarf accretes.
Dwarf novae are a group of cataclysmic variables, showing
repetitive outbursts having amplitudes of 2--8 mag (\cite{war95book}).
In the quiescent state of dwarf novae, the optical emission is a
superposition of the flux from several components,
that is, the thermal emission from the white dwarf and the secondary
star, and free-free emission from an optically thin accretion disk
and a hot spot where the gas stream from the secondary hits the disk
(\cite{szkody1976freefree-temp}).
In an outburst, the thermal emission from an optically thick disk is
dominant (\cite{clarke1984spectrum}; \cite{horne1990spectral}).
SU~UMa-type dwarf novae are a subgroup of dwarf novae, exhibiting two 
types of outburst: a short normal outburst and a long
superoutburst.
During superoutbursts, their light curves show short-term periodic
modulations, called ``superhumps'', which have
a period that is a few percent longer than the orbital period
(\cite{war85suuma}).

It is widely accepted that dwarf nova outbursts can be explained by two
types of instabilities in the accretion disk (\cite{osaki1996DI}).
The first instability is a thermal instability
(\cite{hoshi1979heatinstabirity}).
According to the thermal instability model, the accretion disk can take
only two thermally stable states.
One is a low-viscosity, cool disk consisting of neutral hydrogen gas.
The other one is a high-viscosity, hot disk consisting of ionized gas.
The disk with partially ionized gas is predicted to be thermally unstable.
Dwarf nova outbursts can be interpreted as a state transition from the
cool to hot state, which occurs when the disk density reaches a critical
value in the cool state.

The second instability is a tidal instability
(\cite{white1988tidleinstabirity}; \cite{osaki1989SU-UMa}).
According to this model, the disk becomes tidally unstable, and deforms
to an eccentric disk when the disk radius reaches the 3:1 resonance
radius.
A strong tidal torque works in an eccentric disk, leading to bright
superoutbursts observed in SU~UMa stars.
An eccentric disk is expected to show prograde precession in the
inertial frame of binary systems.
The superhump period, slightly longer than the orbital period, can
naturally be explained by this precession.

WZ~Sge-type dwarf novae form a subclass of SU~UMa stars, which only
experiences superoutbursts (\cite{odo91wzsge}; \cite{osa95wzsge}).
They are characterized by quite long ($\sim 10$~yr) intervals of
superoutbursts.
WZ~Sge stars have received attention because they show
peculiar variations which are not seen in ordinary SU~UMa stars, and
whose mechanisms are poorly understood.

WZ~Sge stars tend to show echo outbursts after the main superoutburst
(\cite{kat04egcnc}).
According to the thermal--tidal instability model, the amount of
gas in the disk should be minimum just after an superoutburst
(\cite{osaki1989SU-UMa}).
The long duration of the echo outburst phase and its sudden cessation
are, hence, problematic for the disk instability model.
\citet{hameury2000echo-hotspot} proposed that an echo outburst is
caused by an enhanced mass-transfer rate from the secondary.
\citet{patterson2002rebright} reported that the observation of WZ~Sge
supported that scenario.
On the other hand, \citet{osaki2001rebright} proposed that an echo
outburst can be triggered if the disk viscosity remains high just after
the outburst, and the gas is supplied from the outer disk (\cite{kat98super}; \cite{kat04egcnc}).
\citet{hellier2001rebright} suggested that a substantial amount of gas
may be stored between the 3:1 resonance and the tidal-limit radius
in binary systems having an extreme mass ratio 
($M_2/M_1 \lesssim 0.1$), such as WZ~Sge-type dwarf novae.
Observational evidence for such a mass reservoir for echo outbursts
has, however, not been established (\cite{uemura2008echo};
\cite{kat08lateSH}).

WZ~Sge-type dwarf novae exhibit unique short-term periodic modulations
only appearing in a very early phase of superoutbursts.
They are called ``early superhumps'', whose period is in agreement with
the orbital period (\cite{patterson1981WZ-Sge-esh};
\cite{kato1996AL-Com-esh}).
Since the amplitude of the early superhump depends on the inclination
angle of binary systems, it is probably attributed not
to the variation due to viscous heating, but to a geometric effect of the
accretion disk (\cite{kato2002origin-esh}).
It is proposed that a part of the disk is vertically expanded, and has
a non-axisymmetric structure (\cite{kato2002origin-esh};
\cite{osaki2002esh}; \cite{kunze2005-21re}).

V455~And was discovered as a dwarf nova candidate from the Hamburg
Quasar Survey.
Follow-up observations showed an orbital period of 81.09~min and
a spectrum similar to that of WZ~Sge
(\cite{araujo-betancor2004period}; \cite{araujo-betancor2005inclination}).
The first-ever recorded outburst of V455 And was discovered on 2007
September 4\footnote{$\langle$http://ooruri.kusastro.kyoto-u.ac.jp/pipermail/vsnet-alert/2007-September/001152.html$\rangle$}.
As reported in the following section of this paper, V455~And was
actually confirmed to be one of the WZ~Sge stars from observations during
the outburst.
The object became a very bright source, reaching the 8th magnitude in
the $V$ band at maximum.
It, hence, provided a great chance to study WZ~Sge phenomena in detail.
We performed optical--near-infrared multi-band photometric observations
in order to provide the color variation associated with a
superoutburst as well as early and ordinary superhumps.
In this paper, we report on the results of our observation.
We, furthermore, investigate the temporal variation of the temperature and
the size of the emitting region of the disk using the multi-band data.
In section~2, the observation and image reduction are described.
Our observational results are shown in section~3.
The implication of the results is discussed in section~4.
Finally, we summarize our findings in section~5. 

\section{Observations}

\begin{figure}
 \begin{center}
  \FigureFile(7cm,7cm){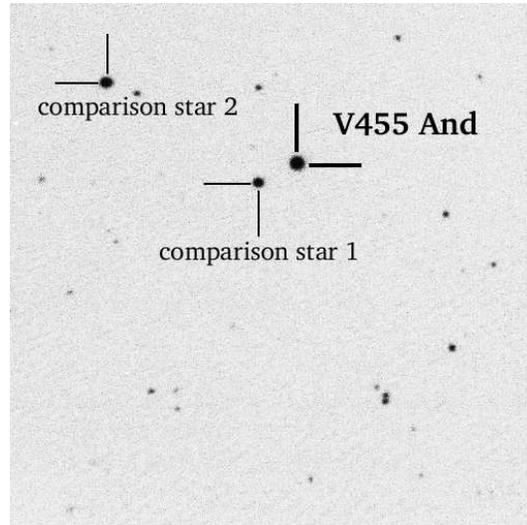}
 \end{center}
 \caption{$V$-band CCD image of the field of V455~And observed on
 15 September 2007 with the
 KANATA telescope. The field of view is 7'$\times$7'.
V455 And and comparison stars are indicated by the black bars.}
\label{fig:v455and}
\end{figure}

We performed photometric observations at two sites.
First, observations at Higashi-Hiroshima Observatory were carried
out with the 1.5-m KANATA telescope.
We used the instrument ``TRISPEC (Triple Range Imager and SPECtrometer
with Polarimetry)'' attached to the Cassegrain focus of
the telescope (\cite{trispec-spec}).
Photometric observations were performed simultaneously in the $V$, $J$,
and $K_s$ bands.
The exposure times of $V$, $J$, and $K_s$-band observations were 10 or
30, 2 or 5, and 1 s, respectively, depending on the sky condition.
An example of the $V$-band images is shown in figure~\ref{fig:v455and}.
Second, the observations at Okayama Astrophysical Observatory were
carried out with the 50-cm MITSuME telescope.
The observations were performed simultaneously in the $g'$,
$Rc$, and $Ic$ bands.
The exposure times of the three bands were between 5 and 60s.
The journal of the observations is
given in tables~\ref{tab:kanata-log} and \ref{tab:mitsume-log} for the
Higashi-Hiroshima and Okayama observatories, respectively.

\begin{table*}
\caption{Observation log for the KANATA telescope}\label{tab:kanata-log}
\begin{center}
  \begin{tabular}{rrrrrr}
     \hline
     T (days) & Time [+JD2454000] & $V$ mag$^*$ & $J$ mag$^*$ & $K_s$ mag$^*$ & Frames \\ 
     \hline
 $-$0.7027---0.6539 & 48.2973---48.3461 & 14.00$\pm$0.03 & 13.09$\pm$0.04 & 12.64$\pm$0.05 & 34 \\
 0.1167---0.3215 & 49.1167---49.3215 & 8.69$\pm$0.01 & 8.82$\pm$0.02 & 8.59$\pm$0.02 & 307 \\
 1.2300---1.3142 & 50.2300---50.3142 & 9.14$\pm$0.02 & 9.27$\pm$0.02 & 8.97$\pm$0.02 & 77 \\
 2.1005---2.2405 & 51.1005---51.2405 & 9.43$\pm$0.02 & 9.50$\pm$0.03 & 9.33$\pm$0.04 & 89 \\
 3.0409---3.1868 & 52.0409---52.1868 & 9.71$\pm$0.01 & 9.75$\pm$0.02 & 9.51$\pm$0.02 & 352 \\
 5.1006---5.2510 & 54.1006---54.2510 & 10.27$\pm$0.01 & 10.27$\pm$0.02 & 10.06$\pm$0.02 & 400 \\
 6.2552---6.3128 & 55.2552---55.3128 & 10.54$\pm$0.01 & 10.48$\pm$0.02 & 10.27$\pm$0.02 & 170 \\
 7.1196---7.2517 & 56.1196---56.2517 & 10.68$\pm$0.01 & 10.59$\pm$0.02 & 10.35$\pm$0.02 & 374 \\
 8.2005---8.2681 & 57.2005---57.2681 & 10.86$\pm$0.01 & 10.75$\pm$0.02 & 10.51$\pm$0.02 & 199 \\
 9.0606---9.3089 & 58.0606---58.3089 & 10.97$\pm$0.01 & 10.84$\pm$0.02 & 10.59$\pm$0.02 & 360 \\
 10.0668---10.2980 & 59.0668---59.2980 & 11.15$\pm$0.01 & 11.02$\pm$0.02 & 10.77$\pm$0.03 & 194 \\
 13.2629---13.3016 & 62.2629---62.3016 & 11.52$\pm$0.01 & 11.37$\pm$0.02 & 11.18$\pm$0.04 & 81 \\
 14.1349---14.3122 & 63.1349---63.3122 & 11.74$\pm$0.01 & 11.53$\pm$0.02 & 11.29$\pm$0.02 & 1048 \\
 15.1447---15.9998 & 64.1447---64.9998 & 11.80$\pm$0.01 & 11.60$\pm$0.02 & 11.34$\pm$0.02 & 790 \\
 16.0002---16.1419 & 65.0002---65.1419 & 11.88$\pm$0.01 & 11.63$\pm$0.02 & 11.35$\pm$0.03 & 466 \\
 16.1350---16.1456 & 65.1350---65.1456 & 12.74$\pm$0.02 & --- & --- & 18 \\
 21.0208---21.0517 & 69.0208---60.0517 & 13.90$\pm$0.01 & 13.11$\pm$0.02 & 12.32$\pm$0.03 & 588 \\
 22.1129---22.2611 & 71.1129---71.2611 & 13.99$\pm$0.02 & 13.09$\pm$0.04 & 12.43$\pm$0.08 & 62 \\
 22.9737---22.9970 & 71.9737---71.9970 & 13.90$\pm$0.02 & 13.02$\pm$0.08 & 12.39$\pm$0.11 & 27 \\
 24.9586---25.3170 & 73.9586---74.3170 & 14.16$\pm$0.02 & 13.30$\pm$0.03 & 12.54$\pm$0.04 & 127 \\
 28.1306---28.3176 & 77.1306---77.3176 & 14.50$\pm$0.01 & 13.68$\pm$0.02 & 12.87$\pm$0.03 & 210 \\
 29.0899---29.2791 & 78.0899---78.2791 & 14.47$\pm$0.02 & 13.62$\pm$0.02 & 12.83$\pm$0.03 & 173 \\
 30.0781---30.2068 & 79.0781---79.2068 & 14.61$\pm$0.01 & 13.80$\pm$0.02 & 13.05$\pm$0.03 & 200 \\
 31.1622---31.1672 & 80.1622---80.1672 & 14.59$\pm$0.02 & 13.77$\pm$0.04 & 13.07$\pm$0.22 & 8 \\
 36.1646---36.2881 & 85.1646---85.2881 & 14.81$\pm$0.02 & 13.95$\pm$0.03 & 13.18$\pm$0.03 & 111 \\
 37.0321---37.3320 & 86.0321---86.3320 & 14.82$\pm$0.02 & 14.00$\pm$0.02 & 13.10$\pm$0.05 & 84 \\
 38.1215---38.1827 & 87.1215---87.1827 & 14.66$\pm$0.01 & 13.96$\pm$0.03 & 12.96$\pm$0.06 & 94 \\
 40.1162---40.2183 & 89.1162---89.2183 & 14.88$\pm$0.01 & 14.09$\pm$0.02 & 13.23$\pm$0.03 & 170 \\
 41.1390---41.2589 & 90.1390---90.2589 & 14.95$\pm$0.01 & 14.13$\pm$0.02 & 13.27$\pm$0.03 & 195 \\
 42.0535---42.1757 & 91.0535---91.1757 & 15.04$\pm$0.01 & 14.23$\pm$0.02 & 13.35$\pm$0.03 & 145 \\
 42.9389---43.2362 & 91.9389---92.2362 & 15.00$\pm$0.02 & 14.25$\pm$0.03 & 13.26$\pm$0.05 & 94 \\
 43.9468---43.9531 & 92.9468---92.9531 & 15.05$\pm$0.02 & 14.19$\pm$0.03 & 13.35$\pm$0.04 & 10 \\
 45.1280---45.1338 & 94.1280---94.1338 & 15.10$\pm$0.02 & 14.22$\pm$0.03 & 13.41$\pm$0.06 & 10 \\
 46.9391---46.9449 & 95.9391---95.9449 & 15.13$\pm$0.03 & 14.43$\pm$0.08 & 12.88$\pm$0.12 & 9 \\
 48.0630---48.2188 & 97.0630---97.2188 & 15.14$\pm$0.01 & 14.35$\pm$0.02 & 13.46$\pm$0.03 & 210 \\
 49.1694---49.1754 & 98.1694---98.1754 & 14.99$\pm$0.02 & 14.19$\pm$0.03 & 13.64$\pm$0.05 & 10 \\
 52.0617---52.1322 & 101.0617---101.1322 & 15.18$\pm$0.01 & 14.41$\pm$0.02 & 13.44$\pm$0.04 & 103 \\
 55.0422---55.1755 & 104.0422---104.1755 & 15.16$\pm$0.01 & 14.35$\pm$0.02 & 13.45$\pm$0.03 & 200 \\
 \hline 
\multicolumn{6}{l}{$^*$Magnitudes are averaged ones in each run.}\\
 \end{tabular}
\end{center}
\end{table*}

\begin{table*}
\caption{Observation log for the MITSuME telescope}\label{tab:mitsume-log}
  \begin{center}
  \begin{tabular}{rrrrrr}
     \hline
   T (days) & Time [+JD2454000] & $g'$ mag$^*$ & $Rc$ mag$^*$ & $Ic$ mag$^*$ & Frames\\
   \hline 
1.9503---2.3055 & 50.9503---51.3055 & 9.33$\pm$0.01 & 9.27$\pm$0.01 & 9.28$\pm$0.01 & 1193\\
5.0391---5.3237 & 54.0391---54.3237 & 10.30$\pm$0.01 & 10.23$\pm$0.01 & 10.19$\pm$0.01 & 2591\\
7.0299---7.3262 & 56.0299---56.3262 & 10.71$\pm$0.01 & 10.61$\pm$0.01 & 10.55$\pm$0.01 & 2700\\
7.9553---8.2677 & 56.9553---57.2677 & 10.91$\pm$0.01 & 10.79$\pm$0.01 & 10.73$\pm$0.01 & 1748\\
8.9281---9.1159 & 57.9281---58.1159 & 11.01$\pm$0.01 & 10.91$\pm$0.01& 10.83$\pm$0.01 & 863\\
12.0480---12.3326 & 61.0480---61.3326 & 11.34$\pm$0.01 & 11.25$\pm$0.01 & 11.17$\pm$0.01 & 683\\
12.9244---13.1539 & 61.9244---62.1539 & 11.50$\pm$0.01 & 11.40$\pm$0.01 & 11.31$\pm$0.01 & 1215\\
13.9231---14.3270 & 62.9231---63.3270 & 11.77$\pm$0.01 & 11.66$\pm$0.01 & --- & 2214\\
15.9210---16.3274 & 64.9210---65.3274 & 11.94$\pm$0.01 & 11.82$\pm$0.01 & 11.70$\pm$0.01 & 1109\\
17.1428---17.2706 & 66.1428---66.2706 & 12.13$\pm$0.01 & 11.97$\pm$0.01 & 11.84$\pm$0.01 & 543\\
19.9168---20.1476 & 68.9168---69.1476 & 13.83$\pm$0.01 & 13.43$\pm$0.01 & 13.20$\pm$0.01 & 983\\
20.9478---21.3074 & 69.9478---60.3074 & 14.07$\pm$0.01 & 13.65$\pm$0.01 & 13.42$\pm$0.01 & 1480\\
21.9398---22.2135 & 70.9398---71.2135 & 14.10$\pm$0.01 & 13.68$\pm$0.01 & 13.45$\pm$0.01 & 1061\\
27.9092---28.3009 & 76.9092---77.3009 & 14.61$\pm$0.01 & 14.26$\pm$0.01 & 14.02$\pm$0.01 & 709\\
29.9078---30.3033 & 78.9078---79.3033 & 14.68$\pm$0.01 & 14.34$\pm$0.01 & 14.12$\pm$0.01 & 831\\
30.9148---31.2400 & 79.9148---70.2400 & 14.70$\pm$0.01 & 14.35$\pm$0.01 & 14.13$\pm$0.01 & 704\\
34.9301---35.0705 & 83.9301---84.0705 & 14.88$\pm$0.01 & 14.51$\pm$0.01 & 14.31$\pm$0.01 & 140\\
38.0765---38.2248 & 87.0765---87.2248 & 14.81$\pm$0.01 & 14.43$\pm$0.01 & 14.22$\pm$0.01 & 224\\
38.8993---39.1974 & 87.8993---88.1974 & 14.90$\pm$0.01 & 14.54$\pm$0.01 & 14.33$\pm$0.01 & 422\\
39.8981---40.2558 & 88.8981---89.2558 & 14.99$\pm$0.01 & 14.66$\pm$0.01 & 14.45$\pm$0.01 & 565\\
40.8971---41.2447 & 89.8971---80.2447 & 15.03$\pm$0.01 & 14.70$\pm$0.01 & 14.50$\pm$0.01 & 510\\
41.8968---42.2413 & 90.8968---91.2413 & 15.12$\pm$0.01 & 14.78$\pm$0.01 & 14.60$\pm$0.01 & 377\\
42.8957---42.9852 & 91.8957---91.9852 & 15.06$\pm$0.02 & 14.71$\pm$0.01 & 14.53$\pm$0.02 & 87\\
43.8952---44.2343 & 92.8952---93.2343 & 15.15$\pm$0.01 & 14.80$\pm$0.01 & 14.61$\pm$0.01 & 352\\
44.9451---45.2134 & 93.9451---94.2134 & 15.16$\pm$0.01 & 14.82$\pm$0.01 & 14.62$\pm$0.01 & 299\\
45.9946---46.2070 & 94.9946---95.2070 & 15.21$\pm$0.01 & 14.88$\pm$0.01 & 14.70$\pm$0.01 & 226\\
46.8927---46.9790 & 95.8927---95.9790 & 15.22$\pm$0.02 & 14.86$\pm$0.01 & 14.69$\pm$0.02 & 99\\
47.9333---48.1951 & 96.9333---97.1951 & 15.21$\pm$0.01 & 14.85$\pm$0.01 & 14.68$\pm$0.01 & 161\\
48.8905---49.1993 & 97.8905---98.1993 & 15.22$\pm$0.01 & 14.87$\pm$0.01 & 14.69$\pm$0.01 & 351\\
52.0067---52.1898 & 101.0067---101.1898 & 15.24$\pm$0.01 & 14.88$\pm$0.01 & 14.71$\pm$0.01 & 152\\
52.8876---53.0824 & 101.8876---102.0824 & 15.20$\pm$0.01 & 14.84$\pm$0.01 & 14.67$\pm$0.01 & 208\\
55.0617---55.1856 & 104.0617---104.1856 & 15.23$\pm$0.01 & 14.89$\pm$0.01 & 14.70$\pm$0.01 & 141\\
\hline
\multicolumn{6}{l}{$^*$Magnitudes are averaged ones in each run.}\\
  \end{tabular}
  \end{center}
\end{table*}

After dark subtraction and flat fielding, we performed aperture
photometry, and obtained differential magnitudes of
the object relative to the comparison stars.
The comparison stars that we used are indicated in figure~\ref{fig:v455and}.
Comparison stars 1 and 2 are located at 
$RA=23^h 34^m04.^s 19$, $DEC=+39\degree 21'24''.1$, and 
$RA=23^h 34^m15.^s 09$, $DEC=+39\degree 22'47''.4$, respectively.
The optical and near-infrared magnitudes of the comparison stars were
quoted from Henden (2006)
\footnote{$\langle$ftp://ftp.aavso.org/public/calib/hs2331.dat$\rangle$}
and the 2MASS catalog \citep{2mass}, respectively.
The magnitude of the comparison star in the $g'$ band was converted
from the $B$
and $V$ magnitudes with the following formula \citep{smith2002g-band}:
\begin{displaymath}
g' = V + 0.54 (B-V) - 0.07.
\end{displaymath}
The magnitudes of the comparison stars are listed in
table~\ref{tab:comp-mag}.

\begin{table}
 \caption{Magnitude of the comparison stars}\label{tab:comp-mag}
 \begin{center}
  \begin{tabular}{lll}
   \hline
    filter & mag (comp.1) & mag (comp.2) \\ 
   \hline
   $g'$  & 13.17 $\pm$  0.01 & 12.45 $\pm$ 0.01\\
   $V$   & 12.70 $\pm$  0.01 & 12.20 $\pm$ 0.01\\
   $Rc$  & 12.12 $\pm$  0.01 & 11.85 $\pm$ 0.01\\
   $Ic$  & 11.63 $\pm$  0.01 & 11.50 $\pm$ 0.02\\
   $J$   & 10.89 $\pm$  0.02 & 11.05 $\pm$ 0.02\\
   $K_s$ & 10.26 $\pm$  0.02 & 10.71 $\pm$ 0.02\\
   \hline 
  \end{tabular}
 \end{center}
\end{table}

The constancy of the magnitude of comparison star 1 was checked with
comparison star 2.
No significant variation was seen over $\sim 0.04$~mag in the relative
magnitude between comparison stars 1 and 2 throughout our
observation.
In the following sections, we show the results using comparison star
1.
We confirmed that the magnitudes calculated with comparison star 2
are in agreement with those with comparison star 1 within the errors in
all bands.
The magnitude errors of the comparison star were included in those of
the object in the following analysis for the spectral energy
distribution (SED).

The interstellar extinction in the direction of V455~And is estimated to
be small, $A_V = 0.34$, according to the database of
\citet{schlegel1998IMR}.
This can be considered as being an upper-limit of the extinction in V455~And.
The actual extinction is probably significantly smaller than the
upper-limit, since V455~And is a nearby source; the distance is
estimated to be 
$90\pm 15\,{\rm pc}$ 
\citep{araujo-betancor2005inclination}.
It is difficult to perform an accurate correction of the extinction
with our available data.
In this paper, we neglect the interstellar extinction.

\section{Result}
\subsection{The 2007 Outburst of V455~And}

\begin{figure}
 \begin{center}
 \FigureFile(9cm,9cm){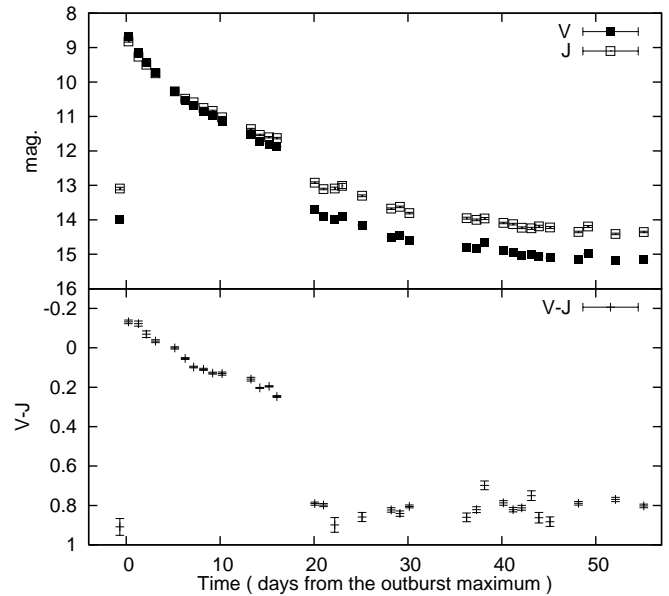}
 \end{center}
 \caption{
 Light curve and color variation during the outburst.
The upper panel shows the light curves in the $V$ and $J$ bands represented
 by the filled and open squares, respectively.
The lower panel shows the color variation of $V-J$.
The figure includes observations for $\sim 2$ months, from 4 September
 to 30 October 2007.
The abscissa is the elapsed days from the outburst maximum (5.5 September
2007(UT) [JD~2454349.0]).  Errors of the photometric points are also
 shown, but they are smaller than the symbol size.
 }
 \label{fig:lc-col}
\end{figure}

Figure~\ref{fig:lc-col} illustrates the overall light curve and color
variation of V455~And during the 2007 outburst.
The outburst of V455~And was discovered on 4 September 2007.
We started to observe the object just after its discovery.
The object was rapidly rising with $-5.8\;{\rm mag}\,{\rm day}^{-1}$ in
the $V$ band during our first night observation on 4 September.
On the next day, 5 September, the outburst reached the maximum when
the magnitude was $8.69\pm 0.01$ in the $V$ band.
Hereafter, we denote the time as $T$, the elapsed days from the outburst
maximum, defining $T=0.0$ as 5.5 September 2007(UT) (JD~2454349.0).

After the object reached the maximum, it had declined from $V=8.7$
to $11.9$ for 17~d.
The fading rate was calculated to be 
$0.31\pm0.02 \;{\rm mag}\,{\rm day}^{-1}$ in the $V$ band at $T=0$--5.
It, then, significantly decreased to 
$0.13 \pm 0.01 \; {\rm mag}\,{\rm day}^{-1}$ at $T=9$--15.
A change of the fading rate occurred at $T\sim 6$.
Such a feature is commonly observed in WZ~Sge-type dwarf novae
\citep{kato2001decline-v}.
The outburst continued at least until $T\sim 17$, and then a rapid
fading from the outburst was observed at $T=20$.
The rapid fading stopped at 2.7 mag brighter than the quiescence in
the $V$ band.
Then, the object again started gradual fading.
Those temporal behaviors are common to the light curves of the other
wave-bands, that is, the $g'$, $Rc$, $Ic$, and $K_s$ bands.
V455~And exhibited no echo outburst, which has been observed
in several WZ~Sge stars \citep{kat04egcnc}.

The color index took a minimum value of $V-J=-0.13\pm 0.01$ when the
brightness was at the maximum.
The color index gradually increased with time during the outburst.
As can be seen in the lower panel of figure~\ref{fig:lc-col}, the color
curve also shows a break at $T\sim 6$.
The color suddenly changed to be red,
$V-J\sim 0.8$, at the same time as when the outburst finished.
After the outburst, the color remained almost constant at $\sim 0.8$,
while the brightness continued a gradual decline.
Thus, the brightness---color relation in
outburst was different from that after the outburst.
This result suggests that a dominant radiation mechanism or component
changed when the outburst finished.

\subsection{SED Variation during the Outburst}

\begin{figure}
 \begin{center}
   \FigureFile(9cm,9cm){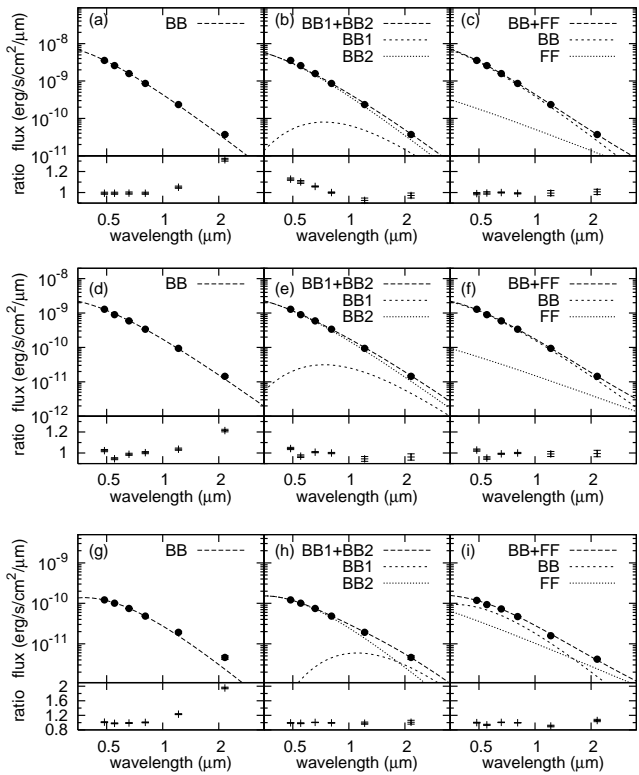}
 \end{center}
 \caption{
Observed SEDs and best models.
The filled circles indicate the observed fluxes in the $g'$, $V$, $Rc$,
 $Ic$, $J$, and $K_s$ bands.
Panels~(a), (b), and (c) show the SEDs at $T=5$ when early
 superhumps appeared.
Panels~(d), (e), and (f) show those at $T=13$ when
 superhump appeared.
Panels~(g), (h), and (i) show those at $T=22$, after
 the outburst.
The model SEDs in the left, middle, and right panels are a blackbody
 radiation, two blackbody components, and the combination model of blackbody and a $10^5$~K
 free-free emissions, respectively.
In the middle and right panels, we show each emission as well as the
 total model SEDs.
Lower frame of each panel shows the ratio of the observed SED to the
 model.  Errors of each point are shown in the figure, but most of them
are smaller than the size of the symbols. 
 }
 \label{fig:sed-day}
\end{figure}

Using 6-band photometric observations, we investigated the radiation
mechanism, the size, and the temperature of the emitting region during
and after the outburst.
Figure~\ref{fig:sed-day} shows the SEDs of the optical--infrared
region.
The top, middle, and bottom panels show the SED at $T=5$, 13, and 22,
respectively.

We need to develop an SED model to obtain physical parameters of the
emitting region for both the outburst and post-outburst states.  It is
considered that the thermal emission from the optically thick disk is
dominant in the optical range during dwarf nova outbursts
(\cite{clarke1984spectrum}; \cite{horne1990spectral}).  It has been
proposed, however, that WZ Sge stars have more complex
emission-components and structure of the accretion disk (\cite{sma93wzsge};
\cite{nog08gwlibv455and}).  In the post-outburst and quiescent states,
furthermore, it has been believed that several components can contribute
to the optical--near-infrared emission, for example, the white dwarf,
the optically thick/thin disk, the hot spot, and the secondary star.  
\citet{araujo-betancor2005inclination} reported that no sign of the
secondary star was detected at quiescence in V455~And.  The
contribution from the secondary star is, hence, negligible in the SED
of V455~And, while the other components are possibly significant.

On the other hand, the ``spectral'' resolution of our 6-band
photometric data is evidently too low to resolve the individual
components.  Our analysis, hence, focused not on the detailed
structure of the emission components, but on the most dominant source
in the optical--near-infrared regime.  As reported in subsection~3.3, early
and ordinary superhumps were observed in all wave-bands during the
outburst of V455~And, which indicates that the optically thick disk is
a dominant source, as in ordinary SU~UMa stars.  
We, hence, tried to fit the SEDs with a blackbody radiation component.
The best-fitted models are indicated by the dashed lines in panels
(a), (d), and (g) in figure~\ref{fig:sed-day}.
The SEDs are well reproduced by the blackbody radiation in the short
wavelength range from the $g'$ to $Ic$ band, as can be seen in the
panels.  On the other hand, there is an excess over
the blackbody emission in the $K_s$-band.
Thus, another emission was dominant in the $K_s$ band in addition to
the blackbody radiation, which was dominant on the blue side.

In order to reproduce the $K_s$-band excess, we attempted to fit the
SEDs using two types of additional components.
The first one was low-temperature blackbody emission.
The obtained best-fitted models are indicated with the dashed line in
panels~(b), (e), and (h) in figure~\ref{fig:sed-day}.
The second model is free-free emission.
In dwarf novae at quiescence, the free-free emission accounts for 
$\sim 50$~\% of the optical flux \citep{szkody1976freefree-temp}.
Its temperature has been estimated to be 
$(0.5$--$1)\times 10^5$~K.
It probably originates from a hot spot or optically thin disk.
Even in outburst, we can expect a contribution of the free-free
emission from the optically thin area in the accretion disk.
The contribution can be large in edge-on systems, such as
V455~And (\cite{araujo-betancor2005inclination}),
because the optically thin area above the optically thick disk is
apparently large.
We tried to fit the SED with a combination model with blackbody and free-free emission.
The temperature of the free-free emission could not, however, be significantly
determined because the dependency of the spectral slope is quite low in
the optical region in the case of a temperature of $\gtrsim 10^4$~K.
The excess over the blackbody emission is, furthermore, too small even
in the near-infrared region to significantly constrain the temperature
of the free-free emission.  Consequently, we fixed the temperature of
the free-free emission as $10^5$~K in all cases.
The best-fitted models are indicated by the dashed lines in panels
(c), (f), and (i) in figure~\ref{fig:sed-day}.
Although the models in the middle and the right panels are different,
both models well reproduce the observed SEDs.
It is difficult to determine the SED model with only our available data.
In this paper, we draw no conclusion about the emission mechanism of
the $K_s$-band excess.  

Table~\ref{tab:temp} gives the temperature of the blackbody emissions
obtained with the three SED models.
The first column gives the time in $T$.
The second column gives the temperature from the single blackbody model.
The third and fourth columns give the temperatures from the
two blackbody model.
The fifth column gives the blackbody temperature from the combination model of blackbody and
free-free emission.
It is preferable that the additional component for the $K_s$-band
excess has little
influence on the high-temperature blackbody component, because the
observed SEDs are almost reproduced by it.
As can be seen from table~\ref{tab:temp}, the temperature makes a sudden
jump at $T=13$ in the two-blackbody model, which is less likely to
indicate a real variation.
It should also be noted that the low-temperature blackbody component
has an atypically high temperature only at $T=2$.
On the other hand, the blackbody temperature from the
blackbody$+$free-free model shows a similar variation to that from the
single blackbody model.  These results indicate that the
blackbody$+$free-free model is more suitable for our analysis.  
Hereafter, we use the blackbody$+$free-free model to describe the
observed SED.  As mentioned above, it does not mean that we conclude
the nature of the $K_s$-band excess to be free-free emission.  The
intensity of the free-free emission can be regarded to be just an
indicator of the degree of the $K_s$-band excess.  

\begin{table*}
  \caption{Temperature of the blackbody emission in the three models.}
\label{tab:temp}
  \begin{center}
    \begin{tabular}{rrrrr}
      \hline
 & Single blackbody model & \multicolumn{2}{c}{Two-blackbody model} & Blackbody$+$free-free model\\
 T (days) & Temp. (K) & Temp.1 (K) & Temp.2 (K) & Temp. (K)\\
\hline
2.16024 & 13090 $\pm$ 540 & 14600 $\pm$ \phantom{x}180 & 5840 $\pm$ 140 & 13440 $\pm$ 370\\
5.18139 & 12560 $\pm$ 380 & 14510 $\pm$ \phantom{x}140 & 3850 $\pm$  \phantom{x}70 & 13300 $\pm$ 330\\
7.17812 & 11730 $\pm$ 320 & 13410 $\pm$ \phantom{x}120 & 3700 $\pm$  \phantom{x}70 & 12360 $\pm$ 340\\
8.13689 & 11550 $\pm$ 310 & 13140 $\pm$ \phantom{x}120 & 3650 $\pm$  \phantom{x}70 & 12200 $\pm$ 180\\
9.00140 & 11510 $\pm$ 370 & 13040 $\pm$ \phantom{x}220 & 3650 $\pm$  \phantom{x}70 & 12080 $\pm$ 160\\
13.0172 & 11110 $\pm$ 370 & 14000 $\pm$ 1060 & 5260 $\pm$ 500 & 11320 $\pm$ 200\\
16.1102 & 10760 $\pm$ 290 & 12710 $\pm$ \phantom{x}800 & 3670 $\pm$ 300 & 11500 $\pm$ 400\\
20.0287 & 7700  $\pm$ 170 &  8280 $\pm$ \phantom{xx}60 & 2890 $\pm$ \phantom{x}50 & 7800 $\pm$ \phantom{x}90\\
21.1294 & 7670  $\pm$ 140 &  8280 $\pm$ \phantom{xx}90 & 2630 $\pm$ \phantom{x}60 & 7780 $\pm$ 190\\
22.0726 & 7620  $\pm$ 170 &  8210 $\pm$ \phantom{xx}70 & 2770 $\pm$ \phantom{x}50 & 7740 $\pm$ \phantom{x}90\\
28.1200 & 7890  $\pm$ 170 &  8900 $\pm$ \phantom{x}320 & 2570 $\pm$ 120 & 8620 $\pm$ 360\\
30.1101 & 7960  $\pm$ 180 &  8820 $\pm$ \phantom{x}220 & 2560 $\pm$ 110 & 8320 $\pm$ 130\\
40.0859 & 8180  $\pm$ 200 &  8620 $\pm$ \phantom{xx}60  & 2700 $\pm$ \phantom{x}20 & 9640 $\pm$ 390\\
41.0600 & 8250  $\pm$ 200 &  9200 $\pm$ \phantom{x}140 & 2470 $\pm$ \phantom{x}50 & 9740 $\pm$ 370\\
44.0659 & 8280  $\pm$ 210 &  9140 $\pm$ \phantom{xx}80  & 2440 $\pm$ \phantom{x}50 & 9520 $\pm$ 260\\
45.0812 & 8150  $\pm$ 200 &  9080 $\pm$ \phantom{x}210 & 2510 $\pm$ \phantom{x}70 & 9320 $\pm$ 400\\
48.0548 & 8100  $\pm$ 190 &  9120 $\pm$ \phantom{xx}60  & 2500 $\pm$ \phantom{x}10 & 8960 $\pm$ 170\\
52.0996 & 8110  $\pm$ 190 &  8970 $\pm$ \phantom{xx}60  & 2440 $\pm$ \phantom{x}50 & 9240 $\pm$ 390\\
55.1236 & 8200  $\pm$ 210 &  8970 $\pm$ \phantom{xx}60  & 2360 $\pm$ \phantom{x}20 & 9400 $\pm$ 310\\
      \hline
  \end{tabular}
  \end{center}
\end{table*}


\begin{figure}
 \begin{center}
   \FigureFile(9cm,9cm){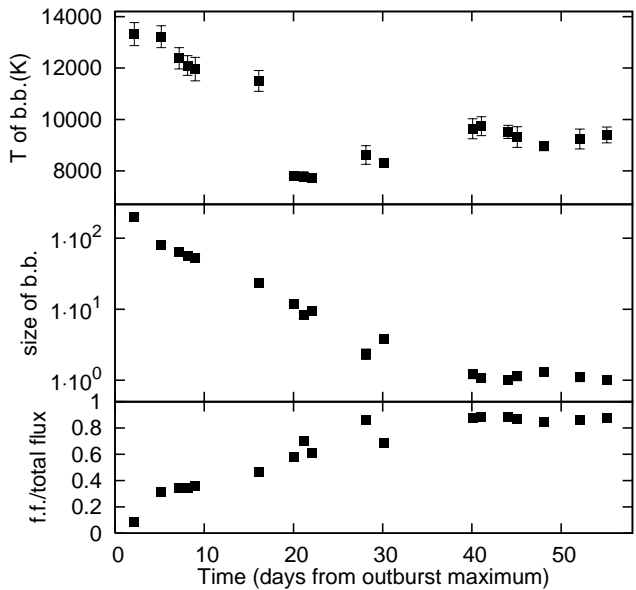}
  \end{center}
  \caption{
Temporal variation of the blackbody temperature (top), size (middle), and
 the ratio of the flux density of the free-free emission to the total
 flux at $2.16\,{\rm \mu m}$.(bottom).
These parameters were estimated with the combination model of a blackbody and
 free-free emission.
The abscissa shows the time in $T$.
The emitting size of the blackbody emission was normalized by the value on
 $T=55$.  Errors are mostly smaller than the symbol size.
}
 \label{fig:temp-day}
\end{figure}

Figure~\ref{fig:temp-day} shows the temporal variation of the best-fitted
parameters of the combination model of blackbody and $10^5$~K
free-free emission.  
The temperature of the blackbody radiation gradually decreases during
the outburst ($T=0$--19).
When the outburst finishes, the temperature rapidly declines to 
$\sim 8000$~K.
It again slightly increases from $T\sim 30$.
It reaches $\sim 10000$~K at $T\sim 40$.
After that, it remains constant.
In contrast to the temperature variation, the emitting size of the
blackbody component gradually decreases during, and even after, the
outburst for $T=0$---$40$.
No dramatic change is seen when the outburst finishes, as can be seen in
the temperature variation.

As shown in the bottom panels of figure~\ref{fig:sed-day}, we found
that the blackbody radiation was still strong on the blue side even
after the outburst.  According to the disk instability model, 
the emission from the optically thick disk should significantly
decrease after the outburst.  It is, hence, not trivial that the
blackbody component just after the outburst originated from the disk.
The disk is, however, still an optimal candidate, because the
temperature of $\sim 8000$~K is 
significantly lower than the white-dwarf temperature in V455~And 
(\cite{araujo-betancor2005inclination})
and much higher than the secondary-star temperature.  
Our result, thus, indicates that a hot optically thick accretion disk
disappeared when the outburst finished, while the disk still remained
optically thick with a moderately high temperature for 10--20~d, even
after the outburst.  The decline of the size stopped at $T\sim 40$, and
subsequently remained constant.
The observed color index of $V-J$ was constant after the outburst,
because $V-J$ of the $\sim 8000$~K blackbody emission was similar to
that of the free-free one.

As shown in the bottom panel of figure~\ref{fig:temp-day}, the
contribution of the free-free emission at $2.16\,{\rm \mu m}$ (or
the degree of the $K_s$-band excess) was small
at the maximum of the outburst.  It, then, gradually increased between
$T=0$--40.  It became dominant after the outburst.
After $T\sim 40$, it was almost constant at a fraction of the total
flux of about 0.85.

\subsection{Short-Term Variation Observed in V455~And}
V455~And showed short-term variations during the outburst.
We report observational features of them in this section.

\begin{figure*}
 \begin{center}
   \FigureFile(18cm,9cm){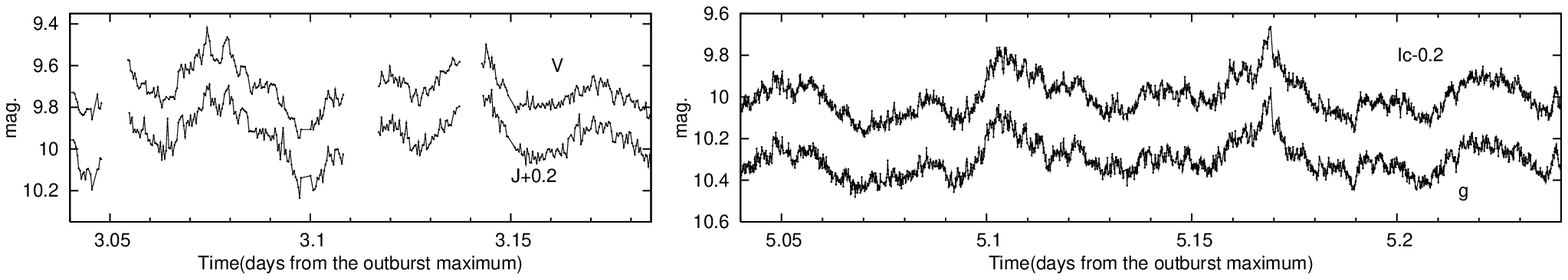}
  \end{center}
 \caption{
 Light curves of early superhumps at $T=3$ and 5.
The abscissa and ordinate denote the time in $T$ and mag, respectively.
The left panel shows the light curves at $T=3$ in $V$ and $J(+0.2)$ mag.
The right panel shows those at $T=5$ in $g'$ and $Ic(-0.2)$ mag.
}
\label{fig:row-esh}
\end{figure*}

Figure~\ref{fig:row-esh} shows the light curves at $T=3$ and 5.
Clear short-term variations were detected with amplitudes of
 $\sim 0.4$~mag on both days.  
Their profile consists of rapid variations with a timescale of 
$\sim 50$~s superimposed on double-peaked periodic modulations (more
clearly seen in the phase-averaged light curves in figure 7).
The double-peaked feature is reminiscent of early superhumps observed in
WZ~Sge-type dwarf nova.

Maehara~et~al. (2007)\footnote{$\langle$http://ooruri.kusastro.kyoto-u.ac.jp/pipermail/vsnet-alert/2007-September/001170.html$\rangle$}
 confirmed that those modulations were actually early superhumps based
 on their period analysis, which showed that the period was in agreement
 with the orbital period of V455~And.
Kato~et~al.
(2007)\footnote{$\langle$http://ooruri.kusastro.kyoto-u.ac.jp/pipermail/vsnet-alert/2007-September/001185.html$\rangle$}
reported the period of the early superhump to be 0.056267$\pm$0.000002~d.


\begin{figure*}
 \begin{center}
   \FigureFile(18cm,9cm){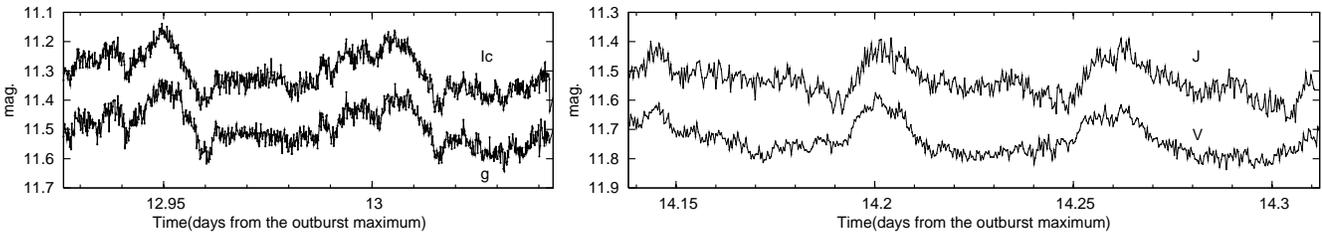}
 \end{center}
 \caption{
 Light curves of superhumps at $T=12$ and 14.
 The abscissa and ordinate denote the time in $T$
 and magnitude, respectively.
 The left panel shows the light curves at $T=12$ in $Ic$ and $g'$ mag.  
 The right panel shows those at  $T=14$ in $V$ and $J$ mag.
 }
\label{fig:row-sh}
\end{figure*}

Figure~\ref{fig:row-sh} shows the short-term variations at $T=12$
 and 14.
In these cases, short-term variations had a single peak profile with
 amplitudes of $\sim 0.2$~mag (also see figure~8).
As in figure~\ref{fig:row-esh}, rapid variations were still seen
particularly at $T=12$.
Their features are consistent with those of superhumps observed in SU~UMa-type
dwarf novae.
Kato~et~al. (2007)\footnote{$\langle$http://ooruri.kusastro.kyoto-u.ac.jp/pipermail/vsnet-alert/2007-September/001204.html$\rangle$}
confirmed that those modulations were actually superhumps based on their
period analysis, which showed that the period was longer than the
orbital period of V455~And.
Maehara~et~al. (2007)\footnote{$\langle$http://ooruri.kusastro.kyoto-u.ac.jp/pipermail/vsnet-alert/2007-September/001222.html$\rangle$}
reported the superhump period to be 0.057093$\pm$0.000015~d.
It is 1.39\% longer than the orbital period of V455~And.

Those detections of the early and ordinary superhumps established the
WZ~Sge-type nature of V455~And.
The outburst in 2007 was, hence, actually a superoutburst of WZ~Sge-type
dwarf novae.
We detected typical early superhumps from $T=0$ to 6.
Photometric studies show that the superhump appeared from $T=7$ and the
early superhump completely disappeared on $T=8$ (Maehara, H. in private
communication).

\subsection{Color and SED Variation Associated with Early and Ordinary
  Superhumps}

Our multicolor photometry allowed us to investigate the color and SED
variations associated with early and ordinary superhumps.
In order to see small color variations, we analyzed phase-averaged
light curves of them for each night.

\begin{figure}
 \begin{center}
   \FigureFile(9cm,9cm){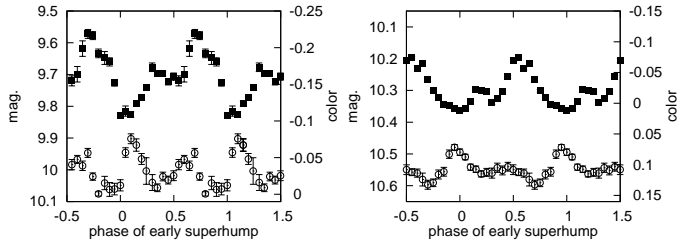}
  \end{center}
 \caption{
Phase-averaged light curves and color variations of early superhumps.
In the left panel, the filled squares and open circles represent the
 $V$-band mag and the $V-J$ color at $T=3$, respectively.
In the right panel, they represent the $g'$-band mag and $g-Ic$ color at
 $T=5$.
The abscissa represents the phase of an early superhump, and its origin shows
 the phase of the early superhump minimum.
}\label{fig:ave-esh}
\end{figure}

Figure~\ref{fig:ave-esh} shows the phase-averaged light curve and the color
variations of early superhumps.
In the left panel, the filled squares and open circles
represent the $V$-band mag and $V-J$ color at $T=3$, respectively.
In the right panel, they represent the $g'$-band mag and
$g-Ic$ color at $T=5$.
The observed light curves were folded with an early superhump period of 
0.056267$\pm$0.000002~d, reported in Kato~et~al. (2007)\footnote{$\langle$http://ooruri.kusastro.kyoto-u.ac.jp/pipermail/vsnet-alert/2007-September/001185.html$\rangle$}.
This is the first time that the color variation was successfully
observed in early superhumps.
The object was the bluest at the early superhump minimum.
This indicates that the temperature of the early superhump light source was
lower than that of an underlying component.
The object became reddest during the fading phase from the maximum to the
minimum of early superhumps.
Those features were commonly seen at $T=3$ and 5.


\begin{figure}
 \begin{center}
   \FigureFile(9cm,9cm){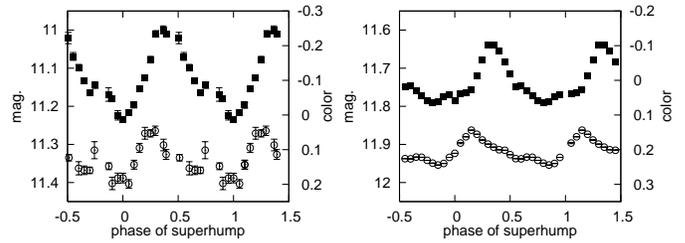}
  \end{center}
 \caption{
 Phase-averaged light curves and color variations of superhumps.
The left and right panels represent the observations at $T=10$ and
 14, respectively.
The filled squares and open circles represent $V$-band mag and
 $V-J$ color, respectively.
The abscissa represents the phase of superhump, and its origin shows the
 phase of the superhump minimum.
} 
 \label{fig:ave-sh}
\end{figure}

Figure~\ref{fig:ave-sh} is the same as figure~\ref{fig:ave-esh}, but for
ordinary superhumps.
The left and right panels represent the observations at $T=10$ and 14,
respectively.
The observed light curves were folded with a superhump period of 
0.057093$\pm$0.000015~d (Maehara~et~al.~2007)\footnote{$\langle$http://ooruri.kusastro.kyoto-u.ac.jp/pipermail/vsnet-alert/2007-September/001222.html$\rangle$}.
The object was reddest at the superhump minimum.
As the brightness increased, the color became bluer.
The object became bluest before the superhump maximum.
Thus, the color behaviors were different from that of early superhumps;
the hump component was red in the early superhump, while it was blue in
the ordinary superhump.


In order to study the temperature variation of the superhump light
source from our data, we should know the structure of emitting
sources.  Early and ordinary superhump light sources have been
considered to be located at an outer part of the accretion disk
(\cite{warnerdonoghue1988hispeed}; \cite{osaki2002esh};
\cite{kato2002origin-esh}).  The inner part of the disk, on the other
hand, probably remains non-variable even during the humps.  The observed
flux can, hence, be a superposition of the light from the hump
component in the outer disk and the non-variable component in the inner
disk.  It might be better that we use an SED model including both the
hump and non-variable components.  As mentioned in subsection~3.2, on the
other hand, it is difficult for our data to resolve both components
without any assumptions.  In this section, we present an analysis
using a simple model, the same as in subsection~3.2: the blackbody and the $10^5$~K
free-free emission model.  In other words, we regard the hump source
as an entire disk in this section.  We also performed an analysis
using the model with both components, while the results depended on  
several assumptions for the non-variable component.  It is presented
in subsection~4.1.

The analysis of SED required simultaneous 6-band photometric data with 
high quality through a few hours of one night.
We had four-night data at $T=5$, 7, 8, and 14 which could be used for our
SED analysis.

\begin{figure}
 \begin{center}
   \FigureFile(9cm,9cm){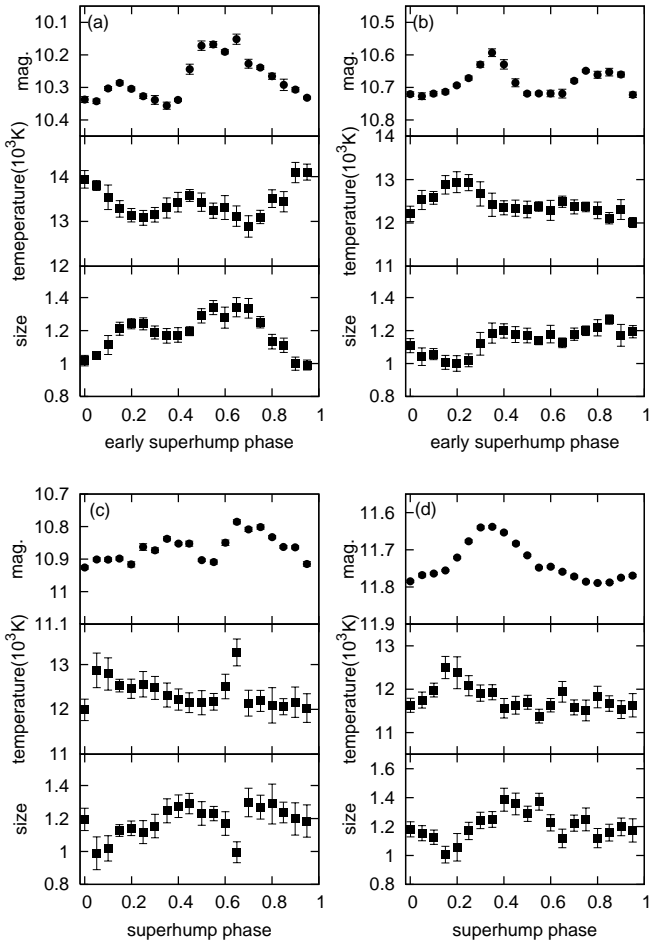}
  \end{center}
 \caption{
Temporal variation of the best-fitted parameters of the blackbody
 emission associated with early and ordinary superhumps.
The top panels show the observed $V$-band light curves.
The middle and bottom panels show the estimated temperature and
the size of blackbody emission region, respectively.
The emitting size was normalized at the phase of the minimum.
The abscissa represents the phase of the humps.
Panels (a), (b), (c), and (d) represent the results at 
$T=5$, 7, 8, and 14, respectively.}
 \label{fig:hump-temp}
\end{figure}

Figure~\ref{fig:hump-temp} shows the temporal variation of the
parameters of our SED model associated with early and ordinary
superhumps.
The top panels show the observed  $V$-band light curves.
The middle and bottom panels show the estimated temperature and the 
size of the blackbody emission region, respectively.
Panels (a), (b), (c), and (d) represent the results
at $T=5$, 7, 8, and 14, respectively.

Figure~\ref{fig:hump-temp}a shows the result for the early superhump.
At the early superhump minimum (phase $\sim 0.00$), the temperature was
highest and the emitting size was smallest.
The light curve correlates well with the emitting size.
This suggests that the nature of early superhumps is the apparent
expansion of a low temperature region.

Figure~\ref{fig:hump-temp}d shows the result for the ordinary
superhump.
From the superhump minimum (phase $\sim 0.00$), the temperature and the
brightness increase, while the emitting size decreases.
The temperature maximum precedes the superhump, and the superhump
reaches maximum after the temperature starts to decline. 
After the superhump maximum, the object fades, first by a decrease
in the temperature, and then by a decrease of the size.  

The accretion disk is believed to be deformed to a precessing eccentric
disk when superhumps are observed.
Due to the orbital motion of the secondary star, the eccentric disk is
periodically heated by a strong tidal torque, making superhumps.
The early phase of the superhump (phase
$\sim 0.00-0.15$) in figure~\ref{fig:hump-temp}d probably corresponds to the phase that the disk temperature increased due to a viscous heating effect.
After the heating stopped, the object entered an expansion--cooling phase.
It was probably an outward expansion of the disk as a result of 
angular-momentum transported from the heated area.

Figure~\ref{fig:hump-temp}b and 9c show short-term modulations during the
transition phase from the early to ordinary superhump phase.
They have small amplitudes and double-peaked profiles.
We can see a hint of two cycles of the heating and expansion-cooling process
associated with the primary and secondary maximum in those panels.
Their behavior is not similar to that of the early superhump, but
rather is similar to that of the ordinary superhump.  The early superhump
signal may have already been quite weak at $T=7$ and 8. 

\begin{figure}
 \begin{center}
   \FigureFile(9.5cm,9cm){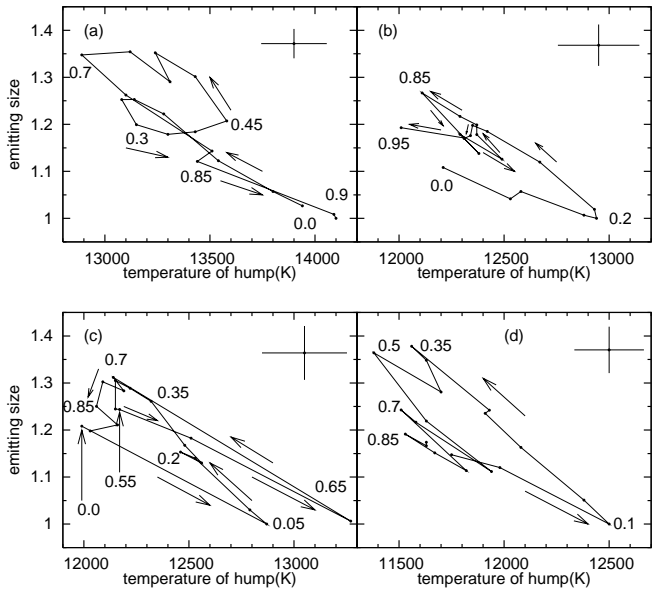}
  \end{center}
 \caption{
Temporal variation of the emitting size against the temperature of the
 blackbody emission.
The abscissa and the ordinate denote the blackbody temperature and the
 emitting size shown in figure~\ref{fig:hump-temp}, respectively.
Typical errors of each point are shown at the upper-right corner of each
 panel.
Panels~(a), (b), (c), and (d) represent the results at $T=5$, 7, 8,
 and 14, respectively.
The numbers represent the hump phase.
} 
 \label{fig:temp-size}
\end{figure}

Figure~\ref{fig:temp-size} shows the temporal variation of the emitting size
against the temperature of the blackbody emission.
Panels~(a), (b), (c), and (d) represent the results
on $T=5$, 7, 8, and 14, respectively.
Panel~(a) shows the result for the early superhump.
First (phase 0.00), the temperature is high, and then gradually declines
as the emission size increase.
Both the emitting size and the temperature increase in phase
$\sim 0.40$.
We note that heating--expanding feature is only seen in panel~(a).
After phase 0.70 in panel~(a), the object returns to the point of
phase 0.00 on a similar track to phase 0.00--0.30.
In the case of the superhump shown in panel~(d), the disk is first
heated and shrunk.
After the temperature reaches the maximum, the object starts rapid cooling and expansion until phase 0.35.
This ``V''-shaped track suggests a process with viscous
heating and cooling by expansion.
Such a ``V''-shaped pattern is not seen in the track of early
superhumps.
In panels~(b) and (c), two cycles of the heating and
cooling process are indicated by the two ``V''-shaped patterns around
phase 0.20 and 0.65 in panel (b) and phase 0.05 and 0.65 in
panel (c).

\section{Discussion}
\subsection{Contribution of the Non-Variable Component from an Inner Part
  of the Accretion Disk}

In subsection~3.4, we analyzed the SEDs, while assuming that the hump
component is the entire disk.  We also analyzed the SEDs with a model
including a 
non-variable component and discuss in this section how it changes the
results. 

\begin{figure}
 \begin{center}
   \FigureFile(8cm,8cm){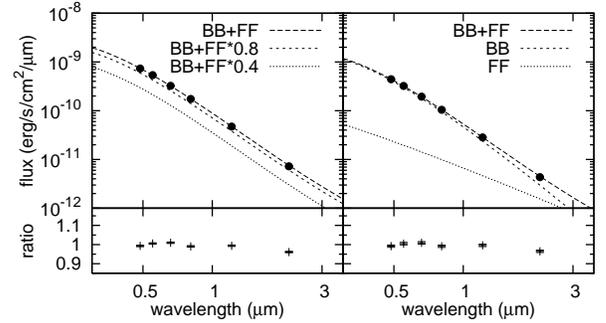}
 \end{center}
 \caption{
Example of the SEDs of the total flux and the hump component at $T=5$ at the
$g'$-band hump minimum.
In the left panel, the filled circles show the observed SED.  
The long dashed, short dashed, and dotted lines denote the best-fitted
 model for the observed SED, 80~\% and 40~\% of the flux of the
 best-fitted model, respectively.
 The latter two models were used as the non-variable component 
 in \S~4.1 (for detail, see the text).
The right panel shows the SED of the hump component in the case that the
 non-variable component is the 40~\% model.
The long dashed, short dashed, and dotted lines denote the best-fitted
 model, its blackbody and free-free components, respectively.
In the upper frames of each panel, the abscissa and ordinate denote the
 wavelength and flux, respectively.
The errors of each point are smaller than the symbols.
The lower frames show the ratio of the observed flux to the model.
}
\label{fig:sed-hump}
\end{figure}

We consider that the observed flux is a sum of the flux from the hump
and the non-variable components.
The SEDs of both components were assumed to be reproduced by the combination model of blackbody and $10^5$~K free-free emission, as used in the last
sections.
We could not, however, significantly determine all parameters of both
components using the 6-band photometric data.
We, hence, make further assumptions for the SED model of the non-variable
component.
The flux ratio of the hump to the non-variable components is, in
particular, difficult to be determined.
The most simple way is to define the flux of the non-variable component as
the observed flux at the hump minimum.
The left panel of figure~\ref{fig:sed-hump} shows the observed SEDs at
the $g'$-band hump minimum at $T=5$.  The long dashed line indicates the
best-fitted model.
If we assume the SED defined by this long dashed line to be a non-variable
component, it means that the flux contribution from the hump source 
is 0~\% at the hump minimum.
It is, however, possible that the hump source had a significant
contribution to the observed flux even at the hump minimum.
The long dashed line in the left panel of figure~\ref{fig:sed-hump} is,
thereby, the possible upper-limit of the flux from the non-variable
component.
In the figure, we also show the model with 80\% and 40\% of the flux of
the best-fitted model at the hump minimum, as indicated by the short
dashed and the dotted lines.
We used the 80 and 40~\% models as the non-variable component in our
analysis.
The right panel of figure~\ref{fig:sed-hump} shows the SED
of the hump component calculated by the subtraction of the SED of the
 non-variable component (the 40~\% model; the dotted line in the left
 panel) from the observed SED. 
As can be seen from the figure, the model well reproduces the SED of
the hump component.
Under those assumptions, we finally obtained the temperature and the
emitting size of the blackbody emission region of the hump component.

\begin{figure}
 \begin{center}
   \FigureFile(9cm,9cm){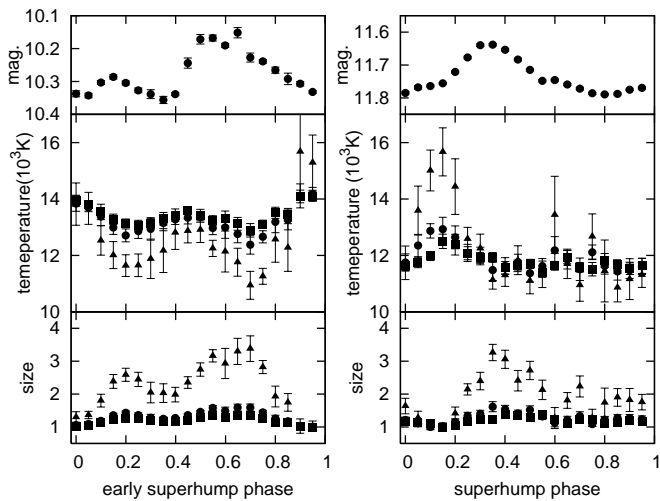}
 \end{center}
 \caption{Temporal variation of the temperature and the size of the
 blackbody emission region from the hump component.
The abscissa represents the phase of the humps.
The top panels show the observed the $V$-band light curve.
The middle and bottom panels show the estimated temperature and the size
 of the emitting region.
The filled squares, circles and triangles represent the results in the
 case that the non-variable component contributes 0\%, 40\%, and 80\% of flux
 at the $g'$-band hump minimum, respectively.
The left and right panels show the results at $T=5$ and $T=14$,
 respectively.
The size of emitting region is normalized at the minimum value.
}
\label{fig:3panel}
\end{figure}

Figure~\ref{fig:3panel} shows the temporal variation of the model
parameters of the blackbody emission for the hump component.
The top, middle, and bottom panels show the same as figure~9.  
The filled squares, circles and triangles in the middle and bottom panels
represent the results in the case that the non-variable component
contributes 0\%, 40\%, and 80\% of the flux at the $g'$-band hump minimum,
respectively.
The result from the 0~\% model corresponds to that reported in subsection~3.4.
The left and right panels show the results of the early superhump at
$T=5$ and the ordinary superhump at $T=14$, respectively.

In the case that the contribution of the non-variable component is large,
the variation amplitudes of the parameters are large.
This can naturally be understood because the observed SED variation
must be reproduced by the flux from a small contribution of the hump
component.
Except for the variation amplitude, the temporal behaviors of the
parameters are qualitatively almost the same in all cases, independent of the
contribution of the non-variable component.

The size of the blackbody emission expands $>3$ times from the hump
minimum in the case of the 80~\% contribution model both in the early
and ordinary superhumps.
Early superhumps are proposed to be variations caused by the vertical
deformation in the disk (\cite{kato2002origin-esh}).
In this case, we can estimate the height, $h$, at the outermost region of the
disk which is required to explain the $>3$ times expansion of the
apparent emitting area.
We consider that we see a flat part ($h=0$) of the disk at an
inclination angle of 75$\degree$ at the hump minimum and a
vertically-expanded part at the hump maximum
(\cite{araujo-betancor2005inclination}).  We calculated the height to be
$h>1.1 r$ at the hump maximum where $r$ is a disk radius.
This height is, however, too large to be explained within the framework of
the standard accretion disk model ($h\sim 0.1r$; \cite{sha73disk}).
The 0\% contribution model yields $h\sim 0.18 r$,
which is still slightly larger than that expected from the standard
model.
These results indicate that the contribution of the non-variable component
in the inner disk should be small in the optical--infrared regime.
In other words, the hump source significantly contributes even at the
hump minimum.

\subsection{Implication of the Remnant Hot Disk in Echo Outbursts}

Some WZ~Sge-type dwarf novae experienced echo outbursts after
the main superoutburst \citep{kat04egcnc}.
The echo outburst can be triggered if a substantial amount of gas is left
at the outermost part of the disk even after the main outburst and works
as a mass reservoir (\cite{kat98super}; \cite{kat04egcnc}; \cite{hellier2001rebright};
\cite{osaki2001rebright}).

V455~And showed no echo outburst, while our analysis revealed an
intriguing feature just after the superoutburst.
As reported in subsection~3.2, the blackbody emission remained at a
moderately high temperature for 10--20 days after the superoutburst.
That period corresponds to the period in which echo outbursts are
observed in the other WZ~Sge stars.
The strong contribution of the blackbody emission indicates the presence
of a luminous accretion disk, in other words, a substantial amount of the
gas remains even after the outburst.
The remnant disk with the moderately high temperature can, therefore, be
a sign of the expected mass reservoir.
V455~And showed no echo outburst, possibly because the amount of the
remnant gas was not enough to trigger an echo outburst.

\citet{hameury2000echo-hotspot} suggests that an echo outburst is caused
by an enhanced mass-transfer rate from the secondary.
In this case, we can expect a high luminosity of the hot spot just after
the outburst, and hence a high contribution of the free-free emission
from the hot spot to the observed SED.
The contribution from the hot spot should then decrease after the echo
outburst phase.
Figure~\ref{fig:temp-day} indicates that the contribution of the
free-free emission keeps a gradual increase for $T=0$--40, and then
becomes 
saturated after $T\sim 40$, without showing the decrease as expected from
the enhanced mass-transfer scenario.  Thus, no sign of the enhanced
mass-transfer rate was confirmed in our data.

After $T\sim40$, the temperature and the size of the blackbody
emission region remains constant.
\citet{araujo-betancor2005inclination} estimates that the surface
temperature of the white dwarf in V455~And is $\sim 10500$~K at
quiescence, which is
close to the estimated temperature after $T\sim40$.
We propose that the blackbody emission after $T\sim40$ originated from the
white dwarf, which becomes a dominant source in the blue part of the SED
due to a weakening of the disk emission.

\subsection{The Origin of the Early Superhump}

Our SED analysis showed that the early superhump light source
is an expanded low-temperature component.  The site and the direction
of the expansion are, however, unclear only in our analysis.  
Early superhumps are considered to be caused by a rotational effect
of the accretion disk in which a part of the disk vertically expands
(\cite{kato2002origin-esh}).  According to the standard disk model,
the temperature of the disk is lower in an outer region.  Our result
in subsection~3.4, hence, indicates that a part of the outer region in the
disk is vertically expanded.  The vertical expansion may be caused 
by scenarios proposed by \citet{kato2002origin-esh} and
\citet{osaki2002esh}.  These models need to be re-examined to explain
the color behavior associated with the early superhumps of V455~And
which were observed for the first time in WZ~Sge stars.

\subsection{Temperature of Superhump: Observations and Theories}

In subsections~3.4 and 4.1, we reported the temperature variation associated
with the superhumps.
There have been only few observations that provide the superhump
temperature based on multi-band photometry.
\citet{hassall1985temp-sh} observed the superoutburst of EK~TrA, and
reported that the superhump light-source has a temperature of 
$\sim 5700$~K.
The color at the superhump maximum was redder than that at the hump
minimum.
\citet{naylor1987sh-temp} reported a similar result for OY~Car,
proposing a superhump temperature of $\sim 8000$~K.
\citet{smak2005temp-sh}, however, theoretically predicted that the
superhump temperature should be quite high in order to reproduce the
observed superhump amplitude.
The superhump temperature is then predicted to be $>15000$~K. 
In this case, the color at the hump maximum should be bluer than that at
the minimum.

Our observation showed that the object reached the hump maximum after
the heating phase ended, and the subsequent expansion started.
\citet{smak2005temp-sh} only discuss the heating phase of superhumps.
The temperature at the hump maximum can, however, be lower than that at
the temperature maximum.
The apparently low-temperature superhumps reported in \citet{hassall1985temp-sh} and
\citet{naylor1987sh-temp} can be explained by a strong contribution of
the expanded low-temperature region.  In \citet{smak2005temp-sh}, an
estimation of the temperature at the superhump light source was
performed with a superhump amplitude of 0.2--0.3~mag.  
In the case of V455~And, the amplitude of the hump was so small ($\sim
0.03$~mag) during the heating phase that the superhump light source
could avoid to take a quite high temperature. 
The discrepancy between the observed and theoretically expected
temperatures can thus be reconciled in our results.

We note, however, that the color behavior associated with
superhumps has not been established.  As also commented in
\citet{smak2005temp-sh}, there are several reports that systems become
bluest at the superhump maximum (\cite{sch80vwhyi};
\cite{sto84tumen}).  Such results are inconsistent with the observed
color behavior in V455~And.  It is possible that WZ~Sge stars have a
different structure of the superhump light source compared with
ordinary SU~UMa stars.  Thus, it is premature to conclude that the
color behavior observed in V455~And is common to all SU~UMa stars. 

\section{Summary}
We performed photometric observations of the 2007 superoutburst of V455~And.
Our 6-band simultaneous observations allowed us to investigate
the temporal variation of the temperature and the size of the emitting region
associated with the superoutburst, early, and ordinary superhumps.
We summarize our findings below.

\begin{itemize}
\item
The optical emission was dominated by blackbody radiation from the accretion disk
     during the superoutburst, while small excesses were found in the
     observed flux in the near-infrared region.

\item
V455~And showed no echo outburst after the superoutburst.
The temperature of the accretion disk rapidly declined at the same
     time when the object entered a rapid fading phase.
The temperature decline then stopped at $\sim 8000$~K.
The size of the emitting region, on the other hand, kept a gradual
     decline.
It indicates that the optically thick disk remained at a moderately high
     temperature even after the superoutburst.
We propose that it is a sign of a mass reservoir that can trigger
     echo outbursts.

\item
A heating and a subsequent expansion-cooling processes were associated
     with the rising phase of superhumps.
The temperature maximum preceded the superhump maximum.
It indicates that the object reached the superhump maximum by the expansion 
of a low temperature region after the heating phase ended.

\item
In a cycle of early superhumps, the object was bluest at the hump
     minimum.
This indicates that the temperature of the early superhump light source
 is lower than that of an underlying component.
The early superhump light source is probably a low-temperature,
     vertically expanded region at the outermost part of the disk.
\end{itemize}

I would like to thank T.Kato and H.Maehara for their valuable comments
on this paper. 
This work was partly supported by a Grant-in-Aid from the Ministry of
Education, Culture, Sports, Science, and Technology of Japan
(19740104).

%

\end{document}